\documentstyle[12pt,fullpage]{article}
\def\be{\begin{equation}}
\def\ee{\end{equation}}
\def\beq{\begin{eqnarray}}
\def\eeq{\end{eqnarray}}
\def\lsim{\:\raisebox{-0.5ex}{$\stackrel{\textstyle<}{\sim}$}\:}
\def\gsim{\:\raisebox{-0.5ex}{$\stackrel{\textstyle>}{\sim}$}\:}
\begin{document}
\begin{flushright}
{\sl TIFR/TH/96-10}
\end{flushright}
\bigskip\bigskip
\begin{center}
{\Large{\bf Like Sign Dilepton Signature for R-Parity Violating
\\[.3cm] SUSY Search at the Tevatron Collider}} \\[2cm]
{\large Monoranjan Guchait\footnote{e-mail: guchait@theory.tifr.res.in}
and D.P. Roy\footnote{e-mail: dproy@theory.tifr.res.in}} \\ 
Theoretical Physics Group \\
Tata Institute of Fundamental Research \\
Homi Bhabha Road, Colaba, Bombay 400 005 \\[4cm]
\underbar{Abstract} \\
\end{center}
\smallskip

The like sign dileptons provide the most promising signature for
superparticle search in a large category of $R$-parity violating SUSY
models.  We estimate the like sign dilepton signals at the Tevatron
collider, predicted by these models, over a wide region of the MSSM
parameter space.  One expects an unambiguous signal upto a gluino mass
of $200 - 300$ GeV ($\geq 500$ GeV) with the present (proposed)
accumulated luminosity of $\sim 0.1~(1)~{\rm fb}^{-1}$.

\newpage

\noindent \underbar{\large{\bf Introduction}} \\
\smallskip

The canonical signature for superparticle search at hadron colliders
is the missing-$p_T$ signature, which follows from $R$-parity
conservation [1].  The latter implies that the superparticles are
produced in pair; and the pair of lightest superparticles (LSP)
resulting from their decays are stable.  In the SUSY models of current
interest the LSP happens to be the lightest neutralino $\tilde Z_1$.
Being a weakly interacting particle, the LSP escapes detection like
the neutrino.  The resulting imbalance of visible transverse-momentum
constitutes the missing-$p_T$ signature for superparticle production.
However, there is a growing interest now in possible signatures for
superparticle production in $R$-violating SUSY models, since there is
no compelling reason for $R$-parity conservation in supersymmetry.  It
is usually invoked to ensure proton stability via the conservation of
lepton and baryon numbers, $L$ and $B$.  However proton stability
requires the conservation of $L$ or $B$, but not necessarily both.
Thus one can have two types of $R$-violating SUSY models,
corresponding to $B$ or $L$ violation, which are consistent with
proton stability.  The former implies LSP decay into a baryonic
channel, which has a large QCD background.  On the other hand the
latter implies LSP decay into a leptonic channel, which can serve as a
viable signature for superparticle search.  In particular the Majorana
nature of the LSP $(\tilde Z_1)$ implies that the LSP pair decay into
like and unlike sign dileptons with equal probability.  This leads to
a like sign dilepton (LSD) signature for superparticle production,
which has very little background from the standard model (SM).

The dilepton data from an early Tevatron run [2], corresponding to an

integrated luminosity of $\sim 4$ pb$^{-1}$, was analysed in [3] in
terms of the above signature.  This led to a lower bound on gluino and
squark masses
\be
m_{\tilde g},m_{\tilde q} \ > \ 100 \ {\rm GeV},
\ee
in the $R$-violating SUSY model.  At present each Tevatron experiment
has accumulated data, corresponding to an integrated luminosity of
$\sim 100$ pb$^{-1}$.  Moreover this is expected to go up by another
order of magnitude following the main injector run (run II).  Thus it
is imperative to explore the prospect of extending the search to
higher superparticle masses in the above mentioned $R$-violating SUSY
model using these data.  The present work is devoted to this exercise.

We shall work within the framework of the minimal supersymmetric
standard model (MSSM), which is characterised by relatively few
parameters [1].  For extending the analysis to $m_{\tilde g} \ > \ 100
\ {\rm GeV}$, one has to include two additional features, which were
not considered in [3].  Firstly one has to consider the cascade decay
of gluino via the chargino and the heavier neutralino states, which
dominate over its direct decay into the LSP for $m_{\tilde g} \geq 100
\ {\rm GeV}$.  Consequently the result will depend not only on the
gluino mass, but on the two other MSSM parameters as well -- i.e. the
higgsino mass parameter $(\mu)$ and the ratio of the two higgs vacuum
expectation values $(\tan\beta)$.  Secondly one has to include the
contribution from the electroweak production of these chargino
$(\tilde W_{1,2})$ and neutralino $(\tilde Z_{1-4})$ states [4].  As we
shall see below, this contribution dominates over the gluino decay
contribution for $m_{\tilde g} \ > \ 200 \ {\rm GeV}$.

In the next section we summarise the $R$-violating decays of LSP,
which will be relevant for our analysis.  In the following section we
shall briefly discuss the processes contributing to LSP production,
i.e. the gluino cascade decay and the electroweak production of the
chargino and neutralino states.  The resulting signal cross-sections
will be presented in the next section.  We shall conclude with a
summary of our results.
\bigskip

\noindent \underbar{\large{\bf LSP Decay in $R$-violating SUSY
Models}}
\smallskip

There can be explicit [5,6] as well as spontaneous [7] breaking of
$R$-parity; but only the former is phenomenologically viable in the
MSSM.  Therefore we shall concentrate on explicit $R$-parity breaking,
where the LSP decay arises from one of the following Yukawa
interaction terms in the Lagrangian:
\be
{\cal L}_{R\!\!\!\!/} = \lambda_{ijk} \ell_i {\tilde\ell}_j \bar e_k +
\lambda'_{ijk} \ell_i \tilde q_j \bar d_k +
\lambda^{\prime\prime}_{ijk} \bar d_i \tilde{\bar d}_j \bar u_k 
\ee
plus analogous terms from the permutation of the supertwiddle.  Here
$\ell$ and $\bar e$ ($q$ and $\bar u,\bar d$) denote the left-handed
lepton doublet and antilepton singlet (quark doublet and antiquark
singlet) and $i,j,k$ are the generation indices.  Evidently the first
two terms violate $L$ and the third one violates $B$ conservation.
The $\lambda$ and $\lambda^{\prime\prime}$ couplings are antisymmetric
in the first two indices, so that there are 9 independent ones of
each.  Together with the 27 $\lambda'$ couplings, there are 45
independent Yukawa coupling terms.  In analogy with the standard higgs
Yukawa couplings, one expects a hierarchical structure for these
additional Yukawa couplings as well [5,6].  The decay channel of LSP
is determined by the leading Yukawa coupling term.

The leptonic decay channels of LSP correspond to one of the $\lambda$
or $\lambda'$ couplings being the leading one, as shown below.
\beq
\lambda'_{3jk} \ &:& \ \tilde Z_1 \rightarrow \tau q \bar q' (\nu_\tau q
\bar q'), \\
\lambda'_{13k,23k} \ &:& \ \tilde Z_1 \rightarrow \nu_\ell b \bar q', \\
\lambda'_{ijk} (i,j \neq 3) \ &:& \ \tilde Z_1 \rightarrow \ell q \bar
q' (\nu_\ell q \bar q'), \\
\lambda_{133,233} \ &:& \ \tilde Z_1 \rightarrow \ell \nu_\tau \bar\tau
(\nu_\ell \tau \bar\tau), \\
\lambda_{123} \ &:& \ \tilde Z_1 \rightarrow \ell \nu_{\ell'} \bar\tau
(\nu_\ell \ell' \bar\tau), \\
\lambda_{311,322,312,321} \ &:& \ \tilde Z_1 \rightarrow \tau \nu_\ell
\bar \ell' (\nu_\tau \ell \bar \ell'), \\
\lambda_{121,122} \ &:& \ \tilde Z_1 \rightarrow \ell \nu_{\ell'} {\bar
\ell}^{\prime\prime} (\nu_\ell \ell' {\bar\ell}^{\prime\prime}),
\eeq
where $\ell$ denotes $e,\mu$ and each of the above final states
represents the corresponding charge conjugate state as well.  Note
that in 15 of these 36 cases, represented by (3) and (4), there is no
$e$ or $\mu$ in the final state.  The $\ell t \bar q'$ final state in
(4) is kinematically inaccessible for the $\tilde Z_1$ mass range of
interest.  Thus the leptonic decay channels of our interest correspond
to the remaining 21 cases, represented by eqs. (5-9).  The
corresponding squared matrix elements are given in [8,9].  For the
$\lambda$ couplings, the pairs of final states shown in eqs. (6-9)
have a branching fraction of $50\%$ each, assuming a common selectron
mass.  For the $\lambda'$ couplings of eq. (5), however, the branching
fraction for the $\tilde Z_1 \rightarrow \ell q \bar q'$ decay is
sensitive to the composition of $\tilde Z_1$.  This is shown in Table
I for different values of the MSSM parameters, where we have assumed a
common sfermion mass $m_{\tilde \ell} = m_{\tilde q} \gg m_{\tilde
Z_1}$. 

It may be noted here that 15 of these 21 cases, represented by
eqs. (5-7), lead to no more than two leptons in the decay of the LSP

pair.  Moreover for the 4 cases represented by eq. (8) the dilepton
final state dominates over the 3-4 lepton state, as we shall see
below.  Thus in 19 of the 21 cases of interest, the LSD channel is the
most viable channel for superparticle search.  Only for the last two
cases, represented by eq. (9), the trilepton final state dominates
over the LSD [10].  Evidently they represent the most favourable case
for the multilepton signal of superparticle production.  We shall not
present the signal cross-section for this case, since it has been
covered in [10].

We shall conservatively assume the leading $R$-violating Yukawa
coupling to be significantly less than 1, so that the pair production
of superparticles and their decays into LSP are not affected.  In this
case the signal does not depend on the value of the Yukawa coupling as
long as it is large enough for LSP decay inside the detector,
i.e. $\gsim 10^{-5}$ [3,6].
\bigskip

\noindent \underbar{\large{\bf Production and Decay of
Superparticles into LSP}}
\smallskip

\nobreak
The MSSM implies $m_{\tilde q} \gsim m_{\tilde g}$.  In estimating
the gluino and chargino/neutralino cross-sections we shall
conservatively assume $m_{\tilde q}$ to be significantly larger than
$m_{\tilde g}$.  In that case the cross-sections are insensitive to
$m_{\tilde q}$ [11].  The results presented below are obtained with
\be
m_{\tilde q} = 2 m_{\tilde g}.
\ee
The dominant processes for gluino production are the leading order
QCD processes [12]
\be
gg (q\bar q) \rightarrow \tilde g \tilde g.
\ee

In order to discuss the cascade decay of gluino, a brief summary of
the chargino/neutralino sector is in order.  The masses of the
$SU(2)$ and $U(1)$ gauginos, $M_2$ and $M_1$, are related to the
gluino mass in the MSSM [1], i.e.
\be
M_2 = {\alpha \over \sin^2 \theta_W \alpha_s} m_{\tilde g} \simeq
0.3 m_{\tilde g},
\ee
\be
M_1 = {5 \over 3} \tan^2 \theta_W M_2 \simeq 0.5 M_2.
\ee
The physical neutralino states $\tilde Z_{1-4}$ are mixtures of
these two gauginos and the two neutral higgsinos.  Similarly the
physical chargino states $\tilde W_{1,2}$ are mixtures of the
charged $SU(2)$ gaugino and the charged higgsino.  Their masses and
compositions are obtained by diagonalising the corresponding mass
matrices [1,13].  They are functions of $m_{\tilde g}$, $\mu$ and
$\tan\beta$ [9,14].  It is important to include the QCD correction
factor, which relates this running gluino mass with its physical
(pole) mass, i.e. [15]
\be
m_{\tilde g} \ {\rm (pole)} \ = m_{\tilde g} (m_{\tilde g}) \left[1
+ {4.2 \alpha_s \over \pi}\right].
\ee
The signal cross-sections are presented below in terms of this pole
mass. 

It is worth mentioning here that for most of the MSSM parameter
space of our interest the higgsino mass parameter $\mu$ is $>
M_1,M_2$.  Consequently the lightest neutralino $(\tilde Z_1)$ is
dominated by the $U(1)$ gaugino component, while the second lightest
neutralino $(\tilde Z_2)$ and the lighter chargino $(\tilde W_1)$
are dominated by the $SU(2)$ gaugino.  Thus their masses roughly
correspond to
\be
m_{\tilde Z_1} \simeq M_1,
\ee
\be
m_{\tilde Z_2} \simeq m_{\tilde W_1} \simeq M_2.
\ee
Moreover, the gluino decays dominantly into these states, i.e.
\beq
\tilde g &\buildrel 0.5 \over \longrightarrow& q \bar q' \tilde W_1,
\\ \tilde g &\buildrel 0.3 \over \longrightarrow& q \bar q \tilde
Z_2, \\
\tilde g &\buildrel 0.2 \over \longrightarrow& q \bar q \tilde Z_1, 
\eeq
where the larger branching fractions into $\tilde W_1$ and $\tilde
Z_2$ reflect the larger $SU(2)$ gauge coupling relative to the
$U(1)$ [9,14].  Of course our results are obtained with exact values
of the masses and branching fractions, which show significant
deviations from the approximate formulae (15-19) over parts of the
parameter space.

In addition to the above, one has to consider the electroweak
processes for chargino/neutralino production [4,16]
\be
q\bar q \rightarrow \tilde W_1 \tilde Z_2, \tilde W_1 \tilde Z_1,
\tilde W_1^+ \tilde W_1^-.
\ee
These are dominated by the $s$-channel $W$ and $Z$ exchanges.  In
spite of being electroweak processes they dominate over the QCD
process (11) for $m_{\tilde g} \ > \ 200 \ {\rm GeV}$ because of the
relatively low $\tilde W_1$ and $\tilde Z_{1,2}$ masses.

Finally the $\tilde W_1$ and $\tilde Z_2$ coming from (17), (18) and
(20) decay into the LSP $(\tilde Z_1)$ via $W$ and $Z$ exchanges,
i.e.
\beq
\tilde W_1 &\buildrel W \over \longrightarrow& q\bar q' \tilde Z_1
(\ell \nu_\ell \tilde Z_1), \\
\tilde Z_2 &\buildrel Z \over \longrightarrow& q \bar q \tilde Z_1
(\ell^+ \ell^- \tilde Z_1).
\eeq
It may be noted that the leptonic decays of $\tilde W_1$ and $\tilde
Z_2$ have branching fractions of 0.22 and 0.06 respectively.  The
former can give rise to a LSD signal for gluino production via (11)
and (17) in the $R$-conserving SUSY model.  Indeed this signal is
expected to be as good as the canonical missing-$p_T$ signal at the
LHC energy [9,14].  For the Tevatron energy, however, the size of
this LSD signal is rather small [17]. On the other hand, one expects
a significant contribution to the LSD signal in the $R$-violating
SUSY model from the cross-term, where one of the leptons comes from
the $\tilde W_1$ (or $\tilde Z_2$) decay.  This contribution is
included in our estimate of the signal cross-section.

The cross-sections presented below are calculated for the Tevatron
collider energy of
\be
\sqrt{s} = 1.8 \ {\rm TeV},
\ee
using the MRSD$_-'$ structure functions [18], with the QCD scale
chosen as the sum of the produced superparticle masses.
\bigskip

\noindent \underbar{\large{\bf Results and Discussion}} \\
\smallskip

\nobreak
We have calculated the signal cross-sections as functions of gluino
mass for 
\be
\tan\beta = 2,10 \ {\rm and} \ \mu = -100,-200,-300,+300 \ {\rm
GeV}.
\ee
A large part of the interval $-100 \ < \ \mu \ < \ 300 \ {\rm GeV}$ is
excluded by the LEP data for the low gluino mass range of our
interest [19], while one expects no significant change in the
cross-sections beyond $|\mu| = 300 \ {\rm GeV}$.

Fig. 1 (a,b,c,d) shows the cross-sections for $\tilde g\tilde g,
\tilde W_1 \tilde Z_2, \tilde W^+_1 \tilde W^-_1$ and $\tilde W_1
\tilde Z_1$ production at $\tan\beta = 2$ and the four values of
$\mu$ mentioned above.  One can clearly see the electro-weak
processes $\tilde W_1 \tilde Z_2$ and $\tilde W_1 \tilde W_1$
overtaking the QCD process $\tilde g\tilde g$ for $m_{\tilde g} \ >
\ 200 \ {\rm GeV}$ at negative values of $\mu$ (Fig. 1a,b,c).  The
rapid increase in the $\tilde W_1 \tilde Z_1$ cross-section as
$m_{\tilde g}$ goes down to $150 \ {\rm GeV}$ reflects production
via on-shell $W$ as $m_{\tilde W_1} + m_{\tilde Z_1}$ falls below
$m_W$.  This enhancement is visible for all the three electroweak
processes at $\mu = 300 \ {\rm GeV}$ (Fig. 1d), as the $\tilde Z_1,
\tilde Z_2$ and $\tilde W_1$ masses are all small for positive
$\mu$.  This is why the region $m_{\tilde g} \lsim 250 \ {\rm GeV}$
is excluded by LEP data for positive $\mu$.  It should be noted,
however, that the electroweak processes dominate over the QCD
process of superparticle production for any gluino mass at positive
$\mu$. 

Fig. 2 (a,b,c,d) shows the cross-sections for the three electroweak
processes at $\tan\beta = 10$.  They are significantly larger than
the previous case at $\mu = -100 \ {\rm GeV}$ (Fig. 2a).  This is
due to a drop in the $\tilde Z_1, \tilde Z_2$ and $\tilde W_1$
masses in going from $\tan\beta = 2$ to $10$ at negative $\mu$,
which is most significant at $\mu = -100 \ {\rm GeV}$.  Note that
all the three cross-sections in Fig. 2a shoot up for $m_{\tilde g}
\simeq 200 \ {\rm GeV}$ due to on-shell $W$ and $Z$ exchanges as the
corresponding thresholds fall below the $W$ and $Z$ masses.  For the
same reason, however, this mass range is ruled out by the LEP data.
On the other hand the cross-sections are clearly smaller than the
previous case at $\mu = +300$ GeV (Fig. 2d).  This reflects an
increase in the above masses in going from $\tan\beta = 2$ to $10$
at positive $\mu$.  Finally the suppression of $\tilde W_1 \tilde
Z_1$ relative to the $\tilde W_1 \tilde Z_2$ and $\tilde W_1 \tilde
W_1$ cross-sections at $\mu = -200,-300$ and $+300 \ {\rm GeV}$ is
due to the decoupling of the $SU(2)$ gaugino component from $\tilde
Z_1$.  All these features can be checked with the masses and
compositions of these particles listed e.g. in [9].

Fortunately the wide variation of the chargino/neutralino
cross-sections with $\mu$ and $\tan\beta$ parameters does not
reflect in the resulting LSD signals.  Lower chargino/neutralino
masses correspond to softer decay leptons; and the resulting
reduction in the detection efficiency compensates for the rise of
the corresponding cross-section.  Moreover the LSP ($\tilde Z_1$)
coming out from on-shell $W$ and $Z$ decays carry very little $p_T$,
so that its decay lepton seldom passes the required $p_T$ cut.  Thus
the sharp peaks seen in some of these cross-sections at low
$m_{\tilde g}$ has little effect on the resulting LSD signals, as we
see below.

The LSD signal coming from the above superparticle decays has been
estimated with the following $p_T$ rapidity and isolation cuts on
each lepton [2]:
\be
p_T^\ell \ > \ 15 \ {\rm GeV}, \ |\eta_\ell| \ < \ 1, \ E_T^{ac} \ < \
5 \ {\rm GeV},
\ee
where the last quantity refers to the transverse energy accompanying
the lepton within a cone of $\Delta R = \sqrt{\Delta \eta^2 + \Delta
\phi^2} = 0.4$.  Our estimates are based on a parton level
Monte-Carlo programme, which should be adequate for the leptonic
signal.  The only place where hadronisation could play a significant
role is in the isolation cut.  We have checked that the efficiency
factors obtained by our parton level programme for the isolation as
well as the $p_T$ and rapidity cuts agree with to those of ISAJET
calculation [20]. 

The LSD background from the standard model has been calculated in
[10].  The total background cross-section is only 2.4 fb, coming
from $WZ$ (2.1 fb) and $t\bar t$ (0.3 fb) [21].  This corresponds to
a quarter of an event for the accumulated luminosity of $\sim 0.1 \
{\rm fb}^{-1}$ in the present Tevatron run, going upto 2.4 events
for the projected luminosity of 1 fb$^{-1}$ with the main injector.
Thus the discovery pottential of the LSD signature is expected to be
limited by the signal size rather than the SM background [10].

Fig. 3(a,b,c,d) shows the LSD signal cross-sections resulting from
$\tilde g\tilde g, \tilde W_1 \tilde Z_2, \tilde W_1^+ \tilde W_1^-$
and $\tilde W_1 \tilde Z_1$ production at $\tan\beta = 2$ and $\mu =
-100,-200,-300$ and $+300 \ {\rm GeV}$, assuming the $\lambda'$
coupling of eq. (5) to be the leading $R$-violating Yukawa coupling.
The relevant branching fractions for $\tilde Z_1 \rightarrow \ell q
\bar q'$ decay are given in Table I.  We see from Fig. 3 that the
LSD signal cross-sections are less sensitive to $\mu$ than the
corresponding raw cross-sections of Fig. 1 as remarked earlier.  The
$\tilde W_1 \tilde Z_2$ and $\tilde W_1^+ \tilde W_1^-$
contributions dominate over $\tilde g\tilde g$ at $m_{\tilde g} \ >
\ 200 \ {\rm GeV}$ through out the negative $\mu$ region, while the
$\tilde W_1 \tilde Z_1$ contribution is very small.  At $\mu = +300
\ {\rm GeV}$, the electroweak contributions dominate over $\tilde
g\tilde g$ for all values of $m_{\tilde g}$.  Note that the size of
the signal cross-section here is quite similar to that in the
negative $\mu$ region.  This is somewhat accidental as the larger
cross-section at $\mu = +300 \ {\rm GeV}$ is compensated by a
smaller branching fraction for the $\tilde Z_1 \rightarrow \ell q
\bar q'$ decay (Table I).

A brief comment on the effect of kimatic cuts is in order.  At
$m_{\tilde g} = 300 \ {\rm GeV}$ and negative $\mu$ the suppression
factors from the lepton $p_T$, rapidity and isolation cuts are
around $5,2$ and $2$ respectively, resulting in an overall
suppression factor of $\sim 20$.  This goes up (down) by a factor of
2 at $m_{\tilde g} = 200 \ (400) \ {\rm GeV}$.  The corresponding
suppression factors at $\mu = +300 \ {\rm GeV}$ are about twice as
large because of the low chargino/neutralino masses.

Fig. 4 shows the net LSD signal cross-sections for different choices of
the leading $R$-violating Yukawa coupling at $\tan\beta = 2$.  Let
us first consider the negative $\mu$ region (Fig. 4a,b,c).  For the
$\lambda'$ and the unfavourable $\lambda$ couplings of eqs. (5) and
(6) one expects a LSD signal cross-section $\geq 100$ fb for
$m_{\tilde g} = 200$ GeV.  For the more favourable $\lambda$
couplings of eqs. (7) and (8) the LSD cross-section remains $\geq
100$ fb upto $m_{\tilde g} = 300$ GeV.  This corresponds to at least
10 isolated LSD events for the current Tevatron luminosity of 0.1
fb$^{-1}$.  Thus the current CDF data is capable of probing for
$R$-violating SUSY signal upto a gluino mass of atleast 200 GeV in
the former case and 300 GeV in the latter.  With the expected
luminosity of $\sim 1 \ {\rm fb}^{-1}$ at the main injector run, the
probe can be extended upto 500 GeV in the former case and $\sim 600$
GeV in the latter.  Turning to the $\mu = +300$ GeV region
(Fig. 4d), one sees that the $\lambda'$ coupling predicts no viable
LSD signal for the current luminosity of 0.1 fb$^{-1}$.  However the
$\lambda$ coupling of eq. (6) predicts a viable signal for
$m_{\tilde g} = 300$ GeV, while for the more favourable couplings of
eqs. (7) and (8) it remains viable upto 500 GeV.  With a luminosity
of 1 fb$^{-1}$ of course one expects a viable signal upto 500 GeV
even for the $\lambda'$ coupling, while for the $\lambda$ couplings
the discovery limit is significantly larger.  It should be noted
here that the four $\lambda$ couplings of eq. (8) can lead to 3 and
4 lepton final states as well.  For this case the trilepton signal,
corresponding to 3 leptons passing the kinematic cut (25), is also
shown in Fig. 4.  It is comparable to the corresponding LSD signal
in the negative $\mu$ region.  But it is relatively small at $\mu =
+300$ GeV, as the low chargino/neutralino masses make it harder for
the 3rd lepton to pass the kinematic cut.  The 4 lepton signal (not
shown) is negligible.

Finally Fig. 5 shows the signal cross-sections for different choices
of the leading $R$-violating Yukawa coupling at $\tan\beta = 10$.
The most noticable feature in this case is the drop in the signal
cross-section for the $\lambda'$ coupling at $\mu = -100$ GeV
(Fig. 5a).  This is due to the fall in the $\tilde Z_1 \rightarrow
\ell q \bar q'$ branching fraction shown in Table I [22].  Apart
from this the signal cross-sections are generally insensitive to the
choice of $\mu$.  With the current luminosity of $\sim 0.1 \ {\rm
fb}^{-1}$, one expects a viable LSD signal for $m_{\tilde g} = 200$
GeV for the $\lambda'$ and the unfavourable $\lambda$ couplings of
eqs. (5) and (6), while for the more favourable $\lambda$ couplings
of eqs. (7) and (8) it remains viable upto $m_{\tilde g} = 400$ GeV.
With a luminosity of $\sim 1$ fb$^{-1}$, the discovery limit goes
upto $m_{\tilde g} = 500$ GeV in the former cases and much beyond in
the latter.  Note that for the $\lambda$ couplings of eq. (8) one
also expects a trilepton signal of similar size.

It may be added here that for the most favourable $\lambda$
couplings of eq. (9), the multilepton signals studied in [10] are
somewhat larger than the ones shown above.  Although these signals
were studied in [10] only for one set of $\mu$ and $\tan\beta$, they
are expected to remain comfortably large for other values of these
parameters as well.
\bigskip

\noindent \underbar{\large{\bf Summary}} \\
\smallskip

\nobreak
The multilepton signals provide the most viable signature for a
large category of $R$-parity violating SUSY models, where the LSP
undergoes leptonic decay via one of the Yukawa couplings of
eqs. (5-9).  In all but 2 of these 21 cases the like sign dileptons
constitute the dominant SUSY signal.  We have done a systematic
analysis of these LSD signals at the Tevatron collider energy
covering a wide range of the MSSM parameters.  The contributions
from the gluino cascade decay as well as the electroweak production
of chargino/neutralino pairs are taken into account.  For the
$\lambda'$ and the unfavourable $\lambda$ couplings of eqs. (5) and
(6) one expects an unambiguous LSD signal upto a gluino mass of at
least 200 GeV with the current accumulated luminosity of $\sim 0.1$
fb$^{-1}$.  For the projected luminosity of $\sim 1$ fb$^{-1}$ from
the main injector run, the signal is expected to remain viable upto
a gluino mass of 500 GeV.  For the more favourable $\lambda$
couplings of eqs. (7) and (8) one expects viable signals upto a
gluino mass of 300 GeV and at least 600 GeV for integrated
luminosities of 0.1 and 1 fb$^{-1}$ respectively.  It may be added
here that the most favourable $\lambda$ couplings of eq. (9), for
which the trilepton signal dominates over the LSD, has been already
investigated in ref [10].  The signal cross-section in this case is
somewhat larger than above.

At present only the CDF experiment can probe for the LSD signal.
But the D$0\!\!\!/$ experiment can also probe for this signal
in the main injector run, since it is scheduled to install a central
magnet for lepton charge identification.  Finally the increase of the
energy from 1.8 to 2 TeV in the main injector run will push up the
corresponding discovery limits somewhat higher. 
\bigskip

\noindent \underbar{\large{\bf Acknowledgements}:} This
investigation was orginally suggested to us by Xerxes Tata as a part
of the low energy SUSY working group activity for the 1994 DPF
Summer Study on High Energy Physics.  We take this opportunity to
express our thanks for the work order along with our apologies for
the late delivery of the goods.  We thank Mike Bisset for helpful
suggestions.

\newpage

\noindent \underbar{\large{\bf References and Footnotes}} \\
\smallskip

\begin{enumerate}
\item[{1.}] For a review see H. Haber and G. Kane, Phys. Rep. 117, 75
(1985). 
\item[{2.}] CDF collaboration : F. Abe et al., Phys. Rev. D45, 3921
(1992). 
\item[{3.}] D.P. Roy, Phys. Lett. B283, 270 (1992).
\item[{4.}] V. Barger, M. Berger, P. Ohmann and R.J.N. Phillips,
Phys. Rev. D50, 4299 (1994).
\item[{5.}] L.J. Hall and M. Suzuki, Nucl. Phys. B231, 419 (1984); 
S. Dawson, Nucl. Phys. B261, 297 (1985); 
S. Dimopoulos and L.J. Hall, Phys. Lett. B207, 210 (1987).
\item[{6.}] H. Dreiner and G.G. Ross, Nucl. Phys. B365, 597 (1991).
\item[{7.}] C.S. Aulakh and R.N. Mahapatra, Phys. Lett. B119, 136
(1983); 
G.G. Ross and J.W.F. Valle, Phys. Lett. B151, 375 (1985).
\item[{8.}] J. Butterworth and H. Dreiner, Nucl. Phys. B397, 3 (1993).
\item[{9.}] H. Dreiner, M. Guchait and D.P. Roy, Phys. Rev. D49, 3270
(1994). 
\item[{10.}] H. Baer, C. Kao and X. Tata, Phys. Rev. D51, 2180 (1995).
\item[{11.}] It is well known that taking $m_{\tilde q}$ comparable to
$m_{\tilde g}$ generally leads to a larger signal cross-section.
However there are some notable exceptions as discussed in [10].
\item[{12.}] E. Reya and D.P. Roy, Phys. Rev. D32, 645 (1985).
\item[{13.}] M. Guchait, Z. Phys. C57, 157 (1993);  Erratum
Z. Phys. C61, 178 (1994).
\item[{14.}] M. Guchait and D.P. Roy, Phys. Rev. D52, 133 (1995).
\item[{15.}] N.V. Krasnikov, Phys. Lett. B345, 25 (1995).
\item[{16.}] H. Baer and X. Tata, Phys. Rev. D47, 2739 (1993); 
H. Baer, C. Chen, C. Kao and X. Tata, Phys. Rev. D52, 1565 (1995).
\item[{17.}] One expects a comparatively larger trilepton signal for
$R$-conserving SUSY model at the Tevatron collider energy via (20-22)
[10]. 
\item[{18.}] A.D. Martin, R.G. Roberts and W.J. Sterling,
Phys. Lett. B306, 145 (1993); B309, 492 (1993).
\item[{19.}] ALEPH collaboration : D. Decamp et al., Phys. Rep. 216,
253 (1992) and Phys. Lett. B349, 238 (1995); L3 collaboration :
O. Adriani et al., Phys. Rep. 236, 1 (1993).
\item[{20.}] F. Paige and S. Protopopescu, in Supercollider Physics,
ed. D. Soper (World Scientific, 1986); H. Baer, F. Paige,
S. Protopopescu and X. Tata, in Proc. of the Workshop on Physics at
Current Accelerators and Supercollider, ed. J. Hewett, A. White and
D. Zeppenfeld (Argonne National Lab. pub. ANL-HEP-CP-93-92, 1993).  We
are grateful to our colleague Nirmalya Parua for doing an ISAJET
calculation of the efficiency factors for one set of MSSM parameters
in order to cross check them with our parton level Monte Carlo
results. 
\item[{21.}] The kinematic cuts used in [10] are $p^\ell_T \ > \ 15 \
{\rm GeV}$, $|\eta_\ell| \ < \ 3$ and $E^{ac}_T \ < \ p^\ell_T/4$.
The isolation cut is essentially the same as ours, while the larger
rapidity interval means a somewhat higher estimate of the LSD
background. 
\item[{22.}] This table shows a continueing fall in the $\tilde Z_1
\rightarrow \ell q \bar q'$ branching fraction as $\tan\beta$ goes up
to 30.  Thus there may be a hole in the parameter space at $\mu =
-100 \ {\rm GeV}$ and large $\tan\beta$, where the LSP decay via the
$\lambda'$ coupling gives no viable LSD signal.
\end{enumerate}

\newpage

\begin{center}
{\large{\bf Table I}}
\end{center}
\begin{enumerate}
\item[{}] The Branching Fraction for LSP decay $(\tilde Z_1
\rightarrow \ell q \bar q')$ via the $R$-violating Yukawa coupling
$\lambda'_{ijk} \ (i,j \neq 3)$
\end{enumerate}
\[
\begin{tabular}{|ccccc|}
\hline
&&&& \\
$m_{\tilde q}$ & $\mu$ & $\tan\beta = 2$ & $\tan\beta = 10$ &
$\tan\beta = 30$ \\
&&&& \\
\hline
&&&& \\
250 & -100 & .87 & .54 & .21 \\
& -200 & .85 & .63 & .46 \\
& -300 & .82 & .61 & .50 \\
& +300 & .15 & .24 & .36 \\
&&&& \\
200 & -100 & .85 & .40 & .18 \\
& -200 & .82 & .57 & .43 \\
& -300 & .77 & .58 & .49 \\
& +300 & .126 & .28 & .38 \\
&&&& \\
300 & -100 & .81 & .27 & .16 \\
& -200 & .77 & .51 & .41 \\
& -300 & .73 & .53 & .47 \\
& +300 & .132 & .33 & .40 \\
&&&& \\
400 & -100 & .76 & .20 & .14 \\
& -200 & .72 & .47 & .39 \\
& -300 & .685 & .51 & .46 \\
& +300 & .166 & .35 & .40 \\
&&&& \\
500 & -100 & .68 & .16 & .13 \\
& -200 & .68 & .45 & .38 \\
& -300 & .65 & .49 & .45 \\
& +300 & .20 & .37 & .41 \\
&&&& \\
\hline
\end{tabular}
\]

\newpage

\noindent \underbar{\large{\bf Figure Captions}} \\
\smallskip

\begin{enumerate}
\item[{Fig. 1.}] The QCD cross-section for gluino pair production
shown along with the electroweak cross-sections for
chargino/neutralino production as functions of gluino mass for
$\tan\beta = 2$ and $\mu = -100,-200,-300$ and $+300 \ {\rm GeV}$.
\item[{Fig. 2.}] Electroweak cross-sections for chargino/neutralino
production shown as functions of gluino mass for $\tan\beta = 10$ and
$\mu = -100,-200,-300$ and $+300 \ {\rm GeV}$.
\item[{Fig. 3.}] The gluino and the chargino/neutralino contributions
to the LSD signal cross-section shown for the $R$-violating Yukawa
couplings of eq. (5) at $\tan\beta = 2$ and different values of
$\mu$. 
\item[{Fig. 4.}] The LSD signal cross-sections shown at $\tan\beta =
2$ and $\mu = -100,-200,-300$ and $+300 \ {\rm GeV}$ for various
$R$-violating Yukawa couplings.  The solid, dot-dashed, long dashed
and short dashed lines correspond to the couplings of eqs. (5), (6),
(7) and (8) respectively.  The trilepton signal cross-section for the
last case is shown as crosses.
\item[{Fig. 5.}] Same as Fig. 4 at $\tan\beta = 10$.
\end{enumerate}
\end{document}